\documentstyle[prl,aps,multicol,epsf]{revtex}
\begin{document}
\draft
\tighten
\title{Absence of Translational Ordering In Driven Vortex Lattices
}
\author{
Steven Spencer$^{\dagger}$ and Henrik Jeldoft Jensen}
\address{Department of Mathematics, Imperial College, 180 Queen's Gate,
London, SW7 2BZ, United Kingdom.}
\date{\today}
\maketitle

\begin{abstract}
By using finite temperature molecular dynamics simulations,
we consider the question of the existence of dynamical ordering as a
two dimensional
vortex lattice is driven through a random background potential.
The shape of the current-voltage($I$-$V$) characteristics are
qualitatively the same as those seen in experiments on the conventional
superconductors $2H$--${\rm NbSe_{2}}$ and ${\rm Mo_{77}Ge_{23}}$.
However, in contrast to previous simulations, we find that the
lattice has no topological or translational
order for any value of the driving but that it does exhibit a
high driving phase where long range orientational order exists.
\end{abstract}


\begin{multicols}{2}
\section{Introduction}
Generic lattices that are driven through a disordered
environment provide a
classic example of an elastic medium that is driven through
a random quenched background potential, and a particular example
of these is the driven vortex lattice in dirty type II superconductors.
Other examples are
charge density waves in disordered conductors, fluids in
porous media and motion of magnetic bubbles through a disordered
medium. In the former case, the vortex lattice, which is set up
when the superconducting material is placed in a magnetic field
of magnitude
between its upper and lower critical fields, is pinned by randomly
positioned defects in the crystalline structure of the material.
When an external current is driven through the material, the motion
of the vortices is also impeded by the disorder in the material.
This results in the existence of a depinning transition
at zero temperature, below which
the average velocity of the vortices (which is proportional to the
voltage across the sample) is zero, while above it is finite.
The depinning transition was initially suggested to be a dynamical
critical phenomenon, with diverging length scales at the depinning
transition, and this was treated for the case of charge density waves
using mean-field elasticity theory \cite{dsf}. The fact that elasticity
theory was used meant that topological defects within the charge density
wave (or vortex lattice) where neglected.

Several years ago, the simulations of Jensen and co-workers 
\cite{hjjnewt,hjjgeilo} showed that as the vortex lattice was driven
through a random quenched background potential, it did not depin as
a single coherently moving elastic medium but, rather, the lattice
began to break and flow plastically. That is to say that there were
regions of the lattice that were pinned by the random potential and then
channels of moving vortices would flow around them. Their simulations
showed that the strength of the random potential that was required to
make the lattice flow plastically vanished in the thermodynamic limit and,
thus, elasticity theory would always break down. Even after taking
this into consideration, elasticity theory has still been frequently
used to study the depinning transition in many systems, particularly the
depinning of a $d-1$ dimensional interface in a $d$-dimensional disordered
environment \cite{leschhorn}.

Recently, the focus has turned away from the study of the transition
itself and turned towards the physics of the moving phase. In particular,
since the vortex lattice flows plastically near to the critical
driving and thus will contain much topological disorder, {\it does the
system exhibit a dynamical phase transition from a low driving phase,
which has a high degree of topological disorder, to a high driving phase
which is topologically ordered?} This question was first addressed by
Shi and Berlinsky \cite{shiberlin} and Koshelev and Vinokur \cite{koshvino}.
Both sets of authors performed computer simulations on simple models
of a two dimensional vortex lattice in a random pinning potential and both
have answered the above question in the affirmative.

Subsequently, Giamarchi and Le Doussal \cite{giamledous} have performed
analytic studies of such a system and they predicted that the high
driving phase of the system is actually a moving {\it topologically
ordered} glass.
Subsequent to this have been further simulations by various authors,
on flux line systems \cite{moon,ryu,marchetti} and on charge density
wave systems \cite{chen}, most of which claim that there is a transition
to a topologically ordered high driving phase.

The experimental situation is much less clear since it is difficult to
examine the fine structure of the moving lattice directly and, thus, indirect
methods can only be used. Experiments on bulk $2H$--${\rm NbSe_{2}}$
\cite{shobo} and
${\rm Mo_{77}Ge_{23}}$ films \cite{hellerqvist} involve measuring the
$I$-$V$ characteristics and then deducing the differential resistance from
the data. These $I$-$V$ curves, which are carried out in the so-called peak
regime \cite{pippard}, bare strong resemblance to early simulations of
Jensen and their shape has been attributed, by the authors, to plastic flow.
In the experiments of Bhattacharya and Higgins \cite{shobo}, they claim that the
dynamical ordering in their experiments occurs when the differential
resistance reaches its high driving asymptotic value, while the authors
of reference \cite{hellerqvist} claim that this occurs at the peak in the
differential resistivity. Since the methods used to make these claims are
indirect, i.e. they are taken from the $I$-$V$ characteristics alone,
simulations are a very useful tool in studying the physics of the driven
system in far more detail. The simulations that we present herein show that
the more correct interpretation is that of reference \cite{shobo}. However,
even though we show that there is a crossover between regimes at the
onset of the
asymptotic flux flow resistance, the high driving phase is not topologically
ordered and topological defects are always frozen into the lattice
structure.

The rest of the paper is outlined as follows: the model and method are
presented, followed by the results, discussion and finally our conclusions.
As per usual.

\section{Model and Method}
The model we consider is that of a two dimensional array of pancake
vortices. The vortices are contained within a rectangular box with
periodic boundary conditions. The dimensions of the box are chosen
such that its size is commensurate with the formation of a perfect 
triangular vortex lattice. That is to say that, for a perfect vortex lattice
of lattice spacing $a_0$ containing $N_{v}=N_{x}N_{y}$ vortices, the
dimensions of the box are $L_{x}=N_{x}a_{0}$ and $L_{y}=N_{y}a_{0}\sqrt{3}/2$,
thus giving an areal density of vortices of $n_{v}=N_{v}/(L_{x}L_{y})=
2/(\sqrt{3}a_{0}^{2})$. In our simulations, we set our unit of length
by $a_{0}=1$ and the system contains $N_{v}=34\times 30=1020$ vortices.
The vortices then interact through a pairwise Gaussian repulsive potential,
such that the potential energy of the interaction between the $ith$ and
the $jth$ vortex is
\begin{equation}
U_{vv}({\bf r_{i}}-{\bf r_{j}})=A_{v}{\rm exp}\left( -\left(
\frac{|{\bf r_{i}}-{\bf r_{j}}|}
{R_{v}}\right)^{2}\right)
\end{equation}
where $A_{v}$ is the strength of the interaction and $R_{v}$ is its range.
We set the energy scale of the simulations here by putting $A_{v}=1$.
In all the simulations presented herein, the vortex-vortex interaction
range is set at $R_{v}=0.6$, which corresponds to a shear modulus of
$c_{66}=0.2695$ and a compression modulus of $c_{11}=1.9943$
\cite{hjjgeilo}.

The pinning potential in the simulations is generated by a set of
$N_{p}$ randomly positioned pinning centres with positions
${\bf R_{j}^{pin}}$. The vortex-pin interaction is then given by
an attractive Gaussian, such that the potential energy of the
interaction between the $ith$ vortex and the $jth$ pin is
\begin{equation}
U_{vp}({\bf r_{i}}-{\bf R_{j}^{pin}})=-A_{p}{\rm exp}\left( -\left(
\frac{|{\bf r_{i}}-{\bf R_{j}^{pin}}|}
{R_{p}}\right)^{2}\right)
\end{equation}
where $A_{p}$ is the strength of the interaction and $R_{p}$ is its range.
The pinning parameters used in these simulations are $R_{p}=0.25$
and $A_{p}=0.5$. The density of pinning centres is $n_{p}=N_{p}/N_{v}=1$.
This set of parameters ensures that the plastic flow of vortices occurs
in the driven system. Let us outline the simulation method below.

From the above model parameters, we can write down a Hamiltonian for
the system
\begin{equation}
{\cal H}= \frac{1}{2}\sum_{i\neq j} U_{vv}({\bf r_{i}}-{\bf r_{j}}) +
\sum_{i=1}^{N_{v}}\sum_{j=1}^{N_{p}}U_{vp}({\bf r_{i}}-{\bf R_{j}^{pin}})
\end{equation}
Before we simulate the $I$--$V$ characteristics of the system, we
relax the vortex configuration into the pinning potential 
at zero driving force. Therefore, the initial conditions for the simulations
of the $I$--$V$ characteristics are a relaxed lattice that already has
many topological defects, rather than a perfectly ordered lattice
as chosen by several other authors \cite{shiberlin,koshvino}.
We do this relaxation by an annealing
technique using Newtonian mechanics and a rescaling of the velocities
after each time step \cite{hjjnewt}. An initial temperature of
$T_{i}=1$ is chosen and then the system is annealed by
slowly reducing the temperature to a final value of $T_{f}=10^{-8}$.
At the end of this part of the simulation, we store the vortex
configuration and use it as the initial configuration for calculating
the dynamical properties of the driven vortex lattice.

From the Hamiltonian above, we can then write down the equation of motion
governing the over-damped dynamics of the driven system
\begin{equation}
\eta\frac{d{\bf r_{i}}}{dt} = -\frac{\partial {\cal H}}{\partial {\bf r_{i}}}
+{\bf F_{dr}}+\chi_{i}(t)
\label{eqofm2}
\end{equation}
the first term on the right hand side of the above equation represents
the total force on vortex number $i$ due to its interaction with all
the other vortices and the pins, the second term is the homogeneous
driving force applied to each vortex ${\bf F_{dr}}=F_{dr}{\bf\hat{x}}$,
which mimics the applied current. These first two terms make up the
deterministic part of the forces, while the third term on the right
hand side is a stochastic term, which has the statistical properties
\begin{equation}
\begin{array}{lll}
\langle\chi_{i}(t)\rangle = 0 & {\rm and} &
\langle\chi_{i}(t)\chi_{j}(t')\rangle = A\delta_{ij}\delta (t-t')
\end{array}
\end{equation}
Here, the angular brackets denote averages over the full distribution
of random variables and
the amplitude of the two-point correlator can be related to an
effective temperature through $A=2k_{B}T\eta$ \cite{hjjalgo}.
In these simulations
we set the Boltzmann constant and the friction coefficient to $k_{B}=\eta=1$
and we use the discrete form for the stochastic forces as outlined in
ref. \cite{hjjalgo}.

Firstly, we perform simulations on the vortex system in the absence of
pinning and at zero driving force $(A_{p}, F_{dr}=0)$. We study
the structure of the lattice
as a function of the temperature to determine the melting scenario for
the system. We measure the density of miscoordinated vortices, $n_{mis}$,
derived
from constructing the Voronoi Diagram for the vortex array \cite{preparata},
the translational order parameter,
\begin{equation}
\Psi_{T} = \left| \left\langle\frac{1}{N_{v}}\sum_{j=1}^{N_{v}}
e^{i{\bf G}\cdot {\bf r_{j}}}\right\rangle\right|^{2}
\end{equation}
and the hexatic order parameter,
\begin{equation}
\Psi_{H} = \left| \left\langle\frac{1}{N_{v}}\sum_{j=1}^{N_{v}}
\frac{1}{z_{j}}\sum_{k=1}^{z_{j}}
e^{i6\theta_{jk}}\right\rangle\right|^{2}
\end{equation}
Here, both sets of angular brackets denote time averages,
${\bf G}$ is a reciprocal lattice vector, $z_{j}$ is the coordination
number of vortex $j$ and $\theta_{jk}$ is the bond angle between
vortex $j$ and its $kth$ neighbour relative to the fixed $x$-direction.
When we calculate these quantities for the driven system, we always
evaluate the quantities in the co-moving frame.

For a two dimensional lattice in the thermodynamic limit, translational
order only exists at $T=0$. For any finite temperature, the long wavelength
modes destroy the translational order and $\Psi_{T}=0$.
However, even though long range order
is absent, the translational correlation function has a power law form
rather than the exponential form of an isotropic liquid.
Thus, the system
is said to have quasi-long
range order.
In continuum elasticity theory,
one can write the mean squared
displacements of the vortices as
\begin{equation}
\langle u^{2}\rangle_{th} =\frac{k_{B}T}{2\pi^{2}}\int_{0}^{k_{BZ}} \frac{d{\bf
k}}
{c_{66}k_{\perp}^{2}}
\end{equation}
where the limit $c_{11}\gg c_{66}$ has been assumed.
Thus, one can see that the divergence at ${\bf k}=0$ causes unbounded
displacements which gives the loss of translational order. One
can prevent this divergence by introducing a lower cut-off
in the integration brought about by considering a finite sized system
of linear extent $L$.
The Brillouin Zone then spans the $k$-vectors $\pi/L <k<\pi/a_{0}$ and one
can perform the integration to give
\begin{equation}
\langle u^{2}\rangle_{th}\approx\frac{k_{B}T}{c_{66}\pi}\ln\frac{L}{a_{0}}
\end{equation}
i.e. bounded displacements. One can then make an estimate of a melting
temperature by applying the Lindemann criterion, i.e. when the displacements
are $u\sim c_{L}a_{0}$ (where $c_{L}$ is the Lindemann number),
the lattice has melted. This gives a melting
temperature, $T_{m}$, of
\begin{equation}
k_{B}T_{m}\approx\frac{c_{66}\pi (c_{L}a_{0})^{2}}{\ln\frac{L}{a_{0}}}
\end{equation}
which vanishes as $L\rightarrow\infty$. One can invert the above equation
to give $L_{or}(T)$. Here, $L_{or}$ is the linear size of a region (in an
infinite system) over which translational order is maintained, and one
can even think of the volume $L_{or}^{2}$ as being analogous to the
correlated volume in collective pinning theory \cite{blattrev},
with the size of the
volume diverging as $T\rightarrow 0$. However, the above equation for
the ``melting'' temperature (written in `` '' because it is really just
a cross-over where $L_{or}$ becomes of the order of the system size)
has only a very weak inverse logarithmic dependence on $L$ and thus we
can expect that for macroscopic, but finite systems, a low temperature
``phase'' exists where translational long range order exists (again
written in `` '' since it is not a true thermodynamic phase transition).

On increasing the temperature, this two dimensional solid then undergoes
a melting into a liquid phase.
In the Kosterlitz-Thouless type
scenario for two dimensional melting, the two dimensional solid with
quasi-long range order melts at a temperature $T_{m}^{KT}=0.62c_{66}
a_{0}^{2}/(4\pi)$ \cite{fishKT}, into a hexatic
liquid (for the system that we consider, this is $T_{m}^{KT}\approx 0.0133$).
This melting is brought about by the unbinding of dislocation pairs
and the hexatic liquid is characterised by short range translational
order (with $\Psi_{T}=0$ as before) and quasi-long range orientational
order, i.e. $\Psi_{H}=0$ with power law hexatic correlations.
At a higher temperature still, the unbound dislocations undergo a
further unbinding by splitting into unbound disclination pairs. Here,
both the translational and orientational order are short ranged and
an isotropic liquid is formed.

\begin{figure}
\narrowtext
\centerline{\epsfxsize=7cm
\epsffile{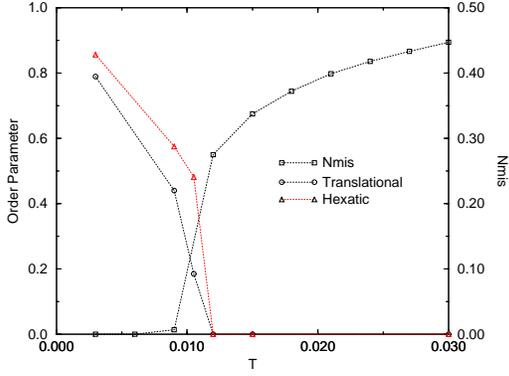}}
\centerline{\epsfxsize=5cm
\epsffile{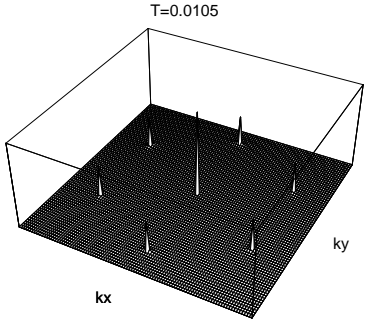}}
\centerline{\epsfxsize=5cm
\epsffile{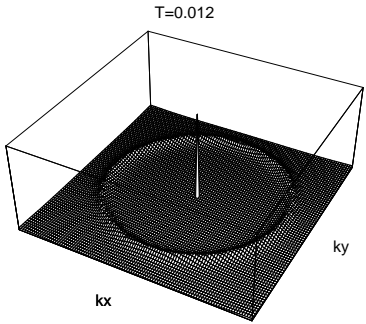}}
\vspace{5pt}
\caption{The pure, undriven system -- the density of miscoordinated vortices
and the translational and hexatic order parameters. Also shown are the
structure factors, $S({\bf k})$, slight below and slightly above the
melting transition.
}
\label{puresys}
\end{figure}

We then simulate the temperature dependence of $n_{mis}, \Psi_{T}$
and $\Psi_{H}$ in order to compare with the above scenario.
Figure \ref{puresys} shows the three quantities along with the structure
factor $S({\bf k})$ for two different temperatures. One can clearly see that
there is a finite temperature, $T_{m}$, at which both the translational and
hexatic order parameters vanish. This also coincides with the rapid
increase in the density of miscoordinated vortices in the system.
The structure factors $S({\bf k})$ are shown for temperatures
just above and below $T_{m}$. The system clearly undergoes a transition
from a solid, signified by the set of sharp delta-function peaks, to an
isotropic liquid, signified by the single $k=0$ peak and concentric
rings. The peaks in the solid with quasi-long range order diverge
algebraically while in the hexatic liquid, $S({\bf k})$ should display
a set of concentric rings with six-fold symmetry \cite{drn_domb,teitel}.
In these simulations, it seems that there is a true solid phase
for low temperatures, i.e. translational and orientational
order exist throughout the considered system ($L_{or}>L$),
which then melts into an isotropic liquid at a finite temperature
$T_{m}$. Thus, no solid with quasi-long range order or hexatic liquid are
observed.

From these simulations, the melting temperature is then $T_{m}\approx 0.011$,
where we have used the structure factors and order parameters as signitures
of the high and low temperature phases.
From these simulations, we can choose an
upper bound for the temperature at which the driven simulations are
to be run, namely for temperatures below $T_{m}$.
This is the obvious
upper bound since, when investigating the dynamical ordering of the vortex
lattice as it is driven through the pinning potential, there can clearly never
be any ordering for temperatures above $T_{m}$.

We now return to the system that contains pinning in order to address
the central question of this article. When driving the lattice through the
random potential,
we solve the set of equations \ref{eqofm2} numerically for a particular
value of the driving force. The numerical time step for the simulations
is chosen as that which gives good energy conservation in the case of
Newtonian dynamics (without velocity rescaling). Obviously energy will
not be totally conserved in this case due to the numerical approximation.
Instead, it has fluctuations around a constant average value.
We chose a time
step of $\delta t=0.01$, which gives a energy fluctuations of about
$0.01\%$ of the average.
The threshold driving force at zero temperature
for the parameters stated is $F_{c}\approx 0.545$. We begin the simulations
at low driving $F_{dr}=0.75$ and from the disordered configuration obtained
after the initial relaxation procedure. The equations of motion are numerically
iterated until the system has had time to respond to the change in driving
and a steady state has been reached. For this condition to be fulfilled,
we find that it is more than sufficient to solve the equations of motion
until the centre of mass position of the
initial vortex configuration has moved 20 lattice spacings. After this
initial period, time averages are performed on the system. Time averages
are performed over a minimum of 10000 time steps and until the averages
have become stable to within a factor of $\pm 10^{-5}$.
This always corresponds to centre
of mass motion of at least 2 lattice spacings, and is usually much more.

\section{Results and Discussion}
Here we present the results of our simulations for the set of system
parameters listed above. The simulations where performed at temperatures
of $T=0.0001$ and $T=0.001$, both below the melting temperature of
$T_{m}\approx 0.011$.

\begin{figure}
{\epsfxsize=18cm
\epsffile{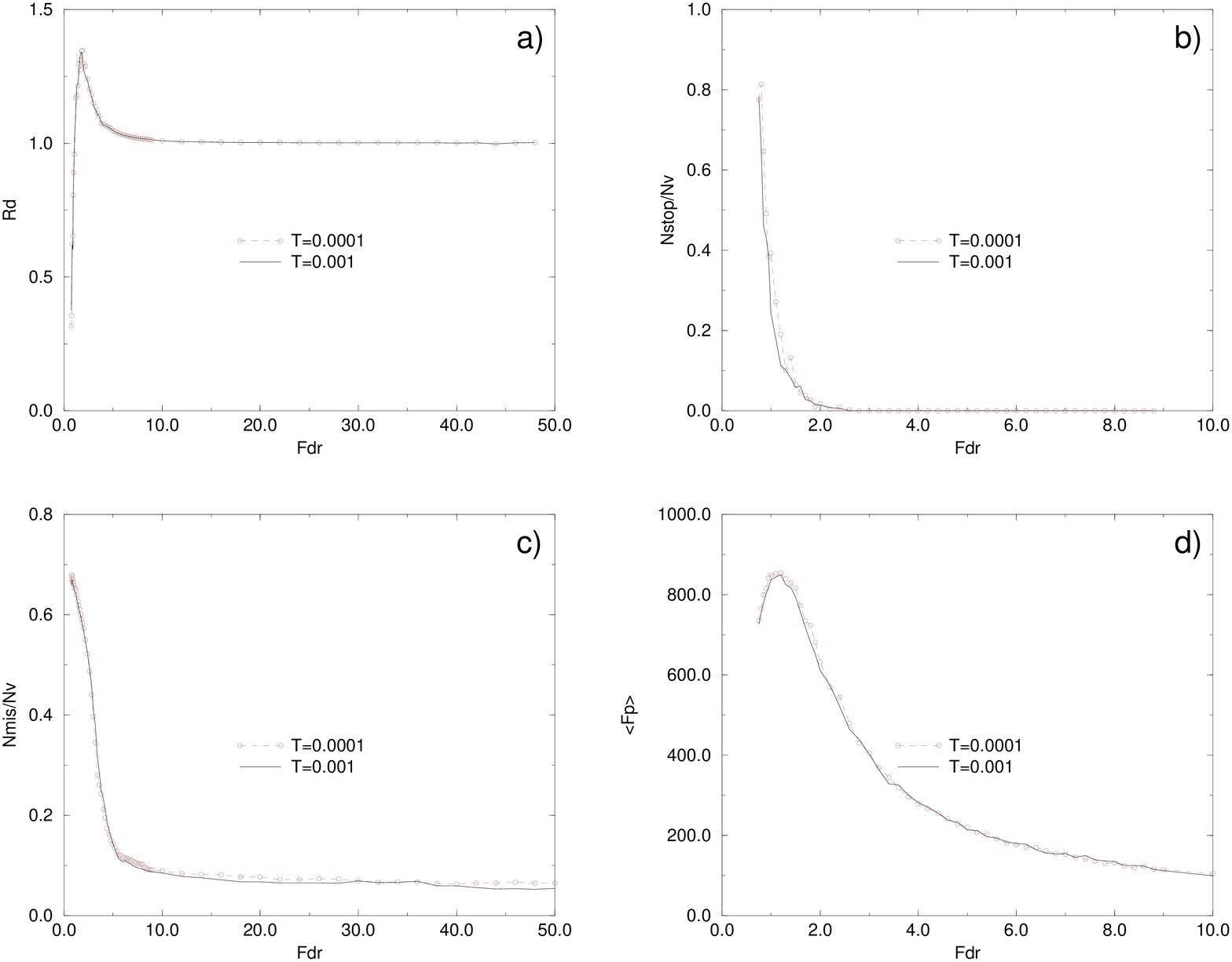}}
\vspace{5pt}
\caption{a) The differential resistance,
b) The fraction of trapped vortices,
c) The fraction of miscoordinated vortices,
d) The average total pinning force. Results for two different temperatures,
$T=0.0001$ and $T=0.001$, are shown.
}
\label{comb_fig}
\end{figure}

The four graphs in Figure \ref{comb_fig} show the differential
resistance $R_{d}=d\langle v\rangle/dF_{dr}$, the fraction of vortices with
average velocity
$\langle v\rangle =0$, $N_{stop}/N_{v}$, the fraction of miscoordinations in the
lattice,
$N_{mis}/N_{v}$,
and the average total pinning force, $F_{p}$, all as a function
of driving force, $F_{dr}$.
\end{multicols}
\widetext
\begin{multicols}{2}

The differential resistance curve is strongly resemblant of that
measured in experimental systems such as $2H$--${\rm NbSe_{2}}$ \cite{shobo}
and those in ${\rm Mo_{77}Ge_{23}}$ thin films \cite{hellerqvist},
where the $I$--$V$
characteristics have
been measured in the peak regime \cite{pippard}.
The peak in this curve, or the fact that
$R_{d}^{max}>R_{d}(F_{dr}\rightarrow\infty )$ is indicative of the
existence of the prominent shoulder feature in the $I$--$V$ curve. 
If one examines the vortex flow patterns at driving forces around the
peak in $R_{d}$, one indeed observes plastic vortex flow in channels,
flowing around regions of pinned vortices.
This is in agreement with the comparison between experimental curves
of this form and this type of vortex flow \cite{shobo}.
Below we present results that show that the only feature of the dynamics of
the system that can be extracted from this particular feature of the $I$--$V$
curve is that plastic flow is occurring in the vortex dynamics. One cannot
determine whether or not dynamical ordering takes place in the system from
this feature alone.

From figure
\ref{comb_fig}a, one can see that the peak in $R_{d}$ occurs at approximately
$F_{dr}\sim 1.8$. From figure \ref{comb_fig}b, one can also see that the
number of trapped vortices vanishes only above a driving of $F_{dr}\sim 2.8$.
From this figure we can easily conclude that, for this particular
system, plastic flow definitely occurs for driving forces $F_{dr}<2.8$. It
should also be noted that this is a lower bound on the threshold for
plastic flow since the true definition of plastic flow is if the
lattice is continually being plastically deformed as it flows, i.e. the
vortices are always flowing plastically if some of the vortices are
constantly changing their neighbours. We can see where the true threshold for
plastic flow is by considering the distributions of the time-averaged 
individual vortex velocities, which are presented below.

Figure \ref{comb_fig}c shows the density of defects as a function of
driving force, with the main features being the strong peak at a driving
force slight greater than threshold and the asymptotic {\it finite} value
of the density for large driving forces. This shows clearly an
absence of a dynamical transition to a topologically ordered
state and this absence is a result of hysteretic effects
in the system. Depending on the initial conditions used in the
simulations, one can obtain the results presented above or one can obtain
the results of various other numerical works, namely that two distinct phases
exist -- a topologically ordered high driving phase and a disordered low driving
phase which may or maybe not separated by a dynamical phase transition.
Now it remains for us to decide which initial conditions are more
viable. In the simulations of Shi and Berlinsky \cite{shiberlin}
and Koshelev and Vinokur \cite{koshvino}, they began their simulations from
an ordered lattice at large driving force. On reducing the driving, they
then see a sharp increase in the density of defects within the vortex array.
Indeed, we can observe this sort of behaviour using the same method.
However, the results we present above correspond to simulations with different
initial conditions. We start with a lattice configuration that is 
initially annealed into the pinning potential and contains a high
degree of topological defects (approximately $50\%$ of the vortices are
miscoordinated). The simulations then start at a driving force slightly
greater than the threshold driving force which is then increased until the
dynamics reach their asymptotic behaviour. We suggest that our initial
conditions are more realistic and correspond to something along the
lines of field-cooling a sample and then measuring the $I$--$V$ curves
by increasing the external current. Thus, in these simulations we see that,
rather than moving into a topologically ordered state on increasing the
driving, the system moves into a state where the number of defects in the
lattice become constant. It is worth noting that in ref. \cite{hellerqvist},
the authors point out that a levelling off of the differential resistance at
high currents is concurrent with a constant density of defects. This is indeed
the case as shown by our simulations, with both quantities reaching their
asymptotic values at around the same value of driving. It is also worth
noting that other recent simulations \cite{marchetti} also observe a
non-vanishing density of defects on increased driving.

The final graph, figure \ref{comb_fig}d, shows the average total pinning
force as a function of $F_{dr}$. Since the physical quantities presented
in figures \ref{comb_fig}a-c are all
a direct consequence of
the existence of the pinning force, the pinning force's
apparent insensitivity to temperature explains the insensitivity to
temperature of the other quantities. We think nobody should
expect that, if the pinning force was temperature independent, strong
temperature dependence should be observed in the other quantities.
For driving forces greater than $F_{dr}\approx 2.0$, the pinning force
is independent of temperature and, in this region, one can fit the behaviour
to $\langle F_{p}\rangle\sim F_{dr}^{-\alpha}$, where the exponent is
$\alpha\approx 9/8$. This compares well with an exponent of unity obtained
from perturbation theory \cite{marchetti}. The fact that
$\langle F_{p}\rangle$ has a power
law dependence may have consequences in the thermodynamic limit, which will
be discussed later.

\begin{figure}
\narrowtext
\centerline{\epsfxsize=7cm
\epsffile{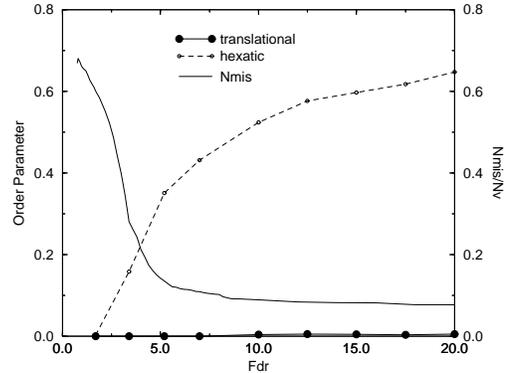}}
\centerline{\epsfxsize=5cm
\epsffile{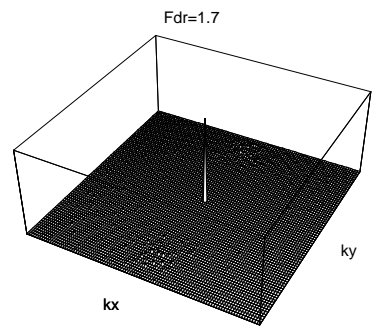}}
\centerline{\epsfxsize=5cm
\epsffile{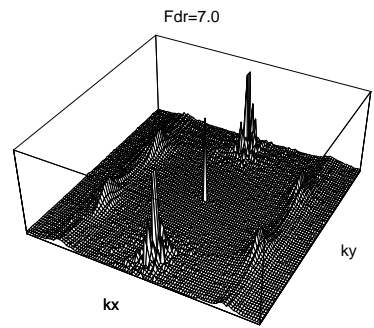}}
\vspace{5pt}
\caption{The same as in figure 1, but for the driven
system. The structure factors are shown for phases with no hexatic
order and for hexatic order only.
}
\label{hexord}
\end{figure}
Thus, we have seen that no topological ordering takes place. This does
not immediately imply there is no order whatsoever within the lattice
as we know from the Kosterlitz-Thouless scenario. Figure \ref{hexord}
shows the density of miscoordinations, the two order parameters
and the structure factors for the
vortex system at a temperature of $T=0.0001$ and for various driving
forces. The translational order parameter, represented by the filled
circles, remains at $\Psi_{T}\approx 0$ for all driving forces, but from
the hexatic order parameter, we can see that, for high driving
forces, the system has some orientational order. This can also
be seen from the two structure factors shown. For $F_{dr}=1.7$,
$S({\bf k})$ just consists of a single peak at $k=0$ with
no other structure. For the large
driving forces, other peaks have appeared. The peaks are anisotropic
corresponding
to the anisotropic nature of the order of the lattice. We are unfortunately
unable to extract the functional form of the decay of the peaks.
Thus, there is a dynamical
ordering that takes place, but it is not to a topologically
ordered state.
Finally we notice that the stucture factor for $F_{dr}=7.0$
is representative of the structure
factors found at any stronger driving force. The peaks do not appear
to become more narrow with increased driving.

\begin{figure}
\centerline{
\epsfxsize=10cm
\epsffile{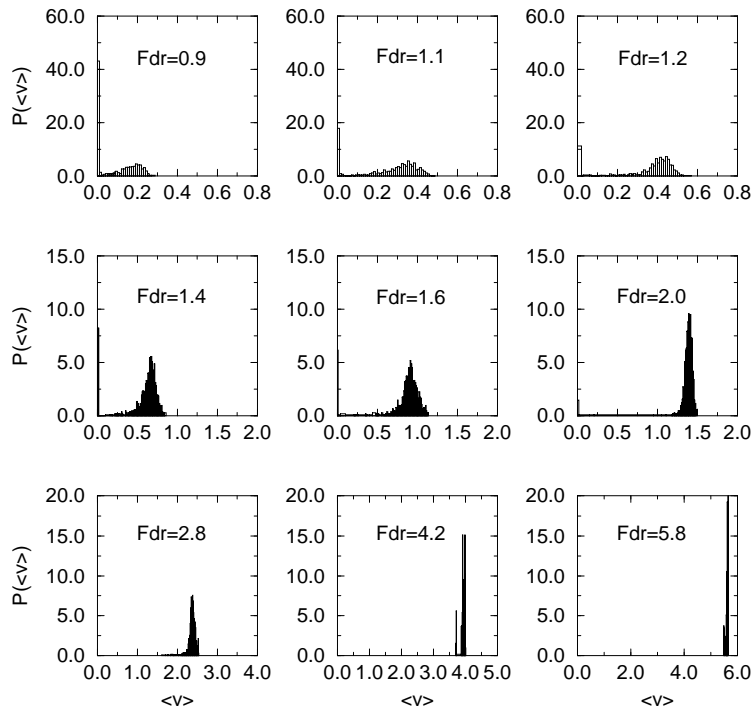}}
\vspace{5pt}
\caption{The distribution of time-averaged velocities
}
\label{pofv}
\end{figure}

Figure \ref{pofv} shows a set of graphs showing the
distribution of the {\it time-averaged}
individual vortex velocities, with each graph corresponding to a different
value of the driving force.
For each of the
$N_{v}$ vortices, we calculate the following
\begin{equation}
\langle v_{i}\rangle =\frac{1}{T}\int_{0}^{T}v_{i}(t)dt
\approx\frac{1}{N\delta t}|{\bf r_{i}(T)}-{\bf r_{i}(0)}|
\end{equation}
i.e. the angular brackets denote a time average and the subscript $i$
denotes the fact that a spatial average has not been performed.
The figures in \ref{pofv} then show the distributions of these quantities.
Looking at the figures, one sees precisely what one would expect for
plastic flow occurring in the system. Slightly above threshold ($F_{dr}=0.9$)
the distribution consists of a strong peak at $\langle v\rangle=0$,
with some small but finite
velocities having lower weights, i.e. most of the vortices are trapped
for this value of the driving, with the few moving ones contributing towards
the average centre of mass velocity of the whole vortex system. The following
graphs for driving $F_{dr}<2.8$ are qualitatively similar, with the peak at
$\langle v\rangle=0$ decreasing and a second peak at finite velocities
developing
as the driving increases. This continues until $F_{dr}\approx 2.8$, where
the peak at $\langle v\rangle=0$ vanishes
and all the vortices become depinned. This
is consistent with the result in figure \ref{comb_fig}b.
The figures corresponding
to $F_{dr}>2.8$ thus consist of a broad distribution with a single maximum.
The position and the height of the maximum of the distribution both increase
correspondingly with the driving force, and the width of the distribution,
which we call $\Delta\langle v\rangle$, clearly decreases on increased driving.
The fact that these distributions are of finite width ($\Delta\langle v\rangle
\neq 0$) is the singular most important fact in determining whether or not
the vortex array is flowing plastically and is therefore of importance when
addressing the question of whereabouts dynamical ordering occurs.

\end{multicols}
\widetext
\begin{multicols}{2}
For the vortex array to undergo a dynamical transition to a topologically
ordered state at some value of the driving
force, all free topological defects within the lattice must vanish and
the lattice must also clearly return to the regime where only
elastic deformations exist. If plastic flow exists then topological
defects will be constantly appearing and disappearing. Thus, as we have
seen from figure \ref{comb_fig}c that topological defects are always present,
we need to tie in all the results presented above to develop an
overall picture of the dynamics. From figure \ref{comb_fig}b we have stated
that, for $F_{dr}<2.8$, some regions of the vortex lattice are permanently
trapped while the other vortices flow aroung them. Thus for driving forces
in the range $F_{c}<F_{dr}<2.8$, plastic flow occurs with some of the
vortices being permanently pinned. This is in contrast to the work of ref.
\cite{marchetti} which states that some of the vortices only remain pinned for
a finite length of time. If one was to plot the {\it instantaneous}
distribution of vortex velocities, one would see the double peaked structure,
similar to ours, but on plotting the distribution of {\it time-averaged}
velocities, the double peaked form may disappear or simply the
$\langle v_{i}\rangle =0$ peak must move to a finite velocity. The fact that
the distributions presented above are time-averaged and show a double
peaked structure with one peak at $\langle v_{i}\rangle =0$ can only lead
to the conclusion that 
these simulations show that the channel
structures that are formed, along which the plastic flow occurs, are
`frozen' in.
This may, of course, not necessarily be a unique structure dependent
only on the 
value of the driving force, but probably will be dependent on the history of the
system.

Therefore, we know that, for
$F_{dr}>2.8$, all of the vortices are depinned, i.e. $\langle v_{i}\rangle >0$
$\forall i$. However, this does not imply that plastic flow has ceased. 
Here we return to the distributions of figure \ref{pofv}. For all plastic
flow in the vortex dynamics to be truly eradicated, the
{\it infinite time-averaged} velocity distribution must be a
delta function, i.e. all velocities are the same and the system
must be in the elastic regime.
Obviously this is
not the case for the time-averages in the simulations. For finite time-averages
the width of the velocity distribution can be finite and still no plastic
flow occurs over the time span. Thus, if the average is performed over a time
$T$ and the width of the velocity distribution is $\Delta\langle v\rangle$,
then one can say that the corresponding {\it width of the
distribution of distances
moved} during the time span of the average has a width of
$\Delta x\approx\Delta\langle v\rangle T$. We can then say that, for no
plastic flow to occur in the simulation, $\Delta x< a_{0}/2$. Thus, say for
a time average performed over 10000 steps (remember $\delta t=0.01$),
plastic flow will always
occur in this time window if the width of the velocity distribution is
$\Delta\langle v\rangle > \Delta\langle v\rangle_{pl} = 0.005$. For
the simulations presented herein, the time-averages are always performed
over a time $T\ge 10000\delta t=100$ and, thus, $\Delta\langle
v\rangle_{pl} = 0.005$ is an upper bound for our simulations, i.e. any
distribution width taken from these simulations that is greater than this
corresponds to plastic flow. Thus we are in a position to tell when then
vortex system has returned to the elastic regime.

From figure \ref{pofv}, one can see that the width of all the velocity
distributions is greater than this threshold for plastic flow. In fact,
only as the driving force $F_{dr}$ approaches a numerical value of
$\approx 10$ does the width $\Delta\langle v\rangle$ become less than
the threshold. Thus, to bring together all the results above, we can
present the following picture.

\begin{figure}
\narrowtext
{\epsfxsize=7cm
\epsffile{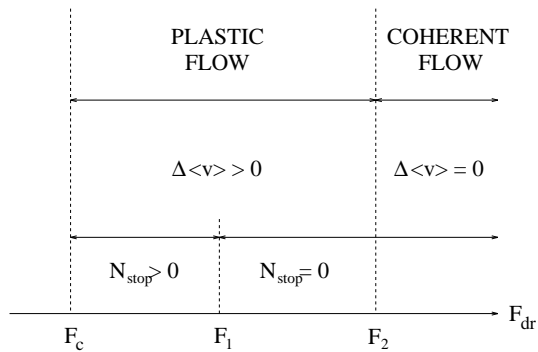}}
\vspace{5pt}
\caption{The dynamical `phases' of the driven vortex system. For the
simulations presented herein, $F_{c}\approx 0.545$, $F_{1}\approx 2.8$
and $F_{2}\approx 10$. See the
text for a more comprehensive explanation.
}
\label{phases}
\end{figure}

The dynamics above threshold can be split into two `phases', with
one of these phases being split into two `sub-phases'. Figure
\ref{phases} shows the scenario that we present as a
schematic diagram. The first phase is for driving forces in the
range $F_{c}<F_{dr}<F_{2}$ ($F_{c}\approx 0.545$, $F_{2}\approx 10$),
where the vortices flow plastically and the
phase can be defined through a non-zero width of the time-averaged individual
vortex velocity distribution (presented in figure \ref{pofv}),
$\Delta\langle v\rangle >0$. This phase is then split into two
sub-phases. These are separated by a threshold $F_{1}\approx 2.8$, which is the
second depinning force. The first depinning force is simply that
which separates phases of zero and non-zero centre of mass velocity,
i.e. $F_{c}$. $F_{1}$ signifies the point where {\it all} of the vortices
become depinned. For $F_{c}<F_{dr}<F_{1}$, some vortices are moving but some
are permanently pinned ($\langle v\rangle_{i} =0$) and thus $N_{stop}>0$.
For $F_{1}<F_{dr}<F_{2}$, all the vortices have finite time-averaged velocities
but the width of the velocity distribution is finite, signifying plastic
flow.

The second phase is for $F_{dr}>F_{2}$ and is defined through
$\Delta\langle v\rangle =0$. Here, all plastic flow has ceased
and the vortex array moves as a single coherent
structure.
That is to say that all the vortices keep the same neighbours
throughout the simulation. Here we use the phrase `single coherent
structure' as opposed to saying that the system has returned to the elastic
regime in order to avoid any association the reader might make between
an elastic regime and a topologically ordered structure.

Given that these are the different regimes of dynamics in these simulations,
we now compare how the time-averaged quantities vary in each of these
regimes. The quantities of interest are the density of miscoordinations,
the two order parameters
and the differential resistance.
Firstly, looking at figure \ref{comb_fig}a, one
can see that the only feature of the differential resistance that correlates
with the regimes described above is that the onset of the
asymptotic flux flow
value of the resistance coincides with the third regime (starting at
$F_{dr}\approx 10$), i.e. where the
plastic flow in the system ceases. A similar thing can be said about the
density of defects. The flattening off of the defect density also coincides
with the onset of this regime. Thus, our main statement of the work
is the following:

There is a lack of a dynamical transition to a topologically ordered
state in two dimensional vortex lattices
with strong pinning. However, there are two possible dynamical regimes
that
can be distinguished. The first is a low driving regime, where the vortices
flow plastically defined through $\Delta\langle v\rangle >0$
(pinned regions may or may not exist).
Here the differential resistance and the density of defects
vary strongly with the driving force. The second regime is the high driving
regime. Here, the vortex array moves as a single coherent structure
($\Delta\langle v\rangle =0$), i.e.
all plastic flow has ceased, and the vortex dynamics have reached their
asymptotic behaviour, i.e. the differential resistance and the density of
defects are independent of the driving force. Since all plastic flow
has ceased in this phase, the defects that remain in the lattice are {\it frozen
in}. This is also reflected in the fact that the translational order
parameter is zero for all driving forces.
Thus, the high driving phase is not a topologically ordered phase,
but one where the vortex lattice is a single coherently moving {\it topological}
glass. However, even though we have shown that no topological or translational
order exist at any driving force, for large enough driving forces the system
exhibits long range orientational order, given by $\Psi_{H}\neq 0$.

\section{Conclusions}

Let us now consider the discrepancy between the simulations in this
article and some recent simulations by other authors
\cite{shiberlin,koshvino,moon,ryu,marchetti}.
The majority of these simulations see
quite a sharp dynamical transition to a topologically ordered
state at some value of the driving force,
which is in complete contrast with the simulations herein.
We suggest that this is either
an artifact of the weak pinning strengths that these simulations have
used or an artifact of the initial conditions, as mentioned
previously (the topological ordering observed by some authors is truly
an {\it artifact} of their initial conditions since they are rather
unrealistic initial conditions).
Here, like most other authors, we are guilty of making a distinction
between weak and strong pinning. In the thermodynamic limit this
distinction becomes meaningless.
In the early simulations of Jensen and co-workers \cite{hjjnewt},
they showed that the pinning strength required to produce plastic flow
in the system vanishes logarithmically as the thermodynamic limit is
approached and, thus, the system is always in the plastic flow regime
\cite{snc}.
We suggest that the same effect could happen here and, thus,
the dynamical ordering seen in other simulations is simply a finite size
effect.

We have also stated here that, on increased driving, our simulations
return to a regime
where plastic flow ceases. Admittedly, this could well be a finite time
effect since we have stated previously that, in the limit of infinite times, the
distribution of time-averaged velocities must be a $\delta$-function for
all plastic flow to cease. Thus, only when the width of the distribution
does vanish will plastic flow truly cease. It is also worth noting that this
may never occur. We have shown that the average pinning force never
vanishes, i.e. it behaves as $\langle F_{p}\rangle\sim 1/F_{dr}$.
This shows that
the asymptotic flux flow resistivity is never actually reached
in this model and,
for infinite systems and times, any finite pinning force is capable of
producing plastic flow.

Thus, we suggest the following. The simulations in this article have still
not completely resolved all the questions as to dynamical ordering in
vortex lattices. What is needed is a detailed finite size study of the
dynamical regimes and topological phases of the lattice so that a 
coherent picture of the dynamical behaviour for the infinite system
can be evaluated. We have shown that, for our system sizes, topological
and translational order do not exist and we believe that considering larger
systems will not change this \cite{hjjnewt}.
However, we have observed a low driving phase
without orientational order and a high driving phase with orientational
order. From a purely theoretical point of view, it would be interesting
to see whether or not these phases persist in the thermodynamic limit and
whether or not they are separated by a true dynamical transition. We have
suggested, above, that plastic flow may persist for all driving forces
in the infinite lattice. From the figures presented herein, we can also see
that the onset of hexatic order overlaps with the region where plastic
flow occurs. Thus, we can suggest that plastic flow does not destroy
the hexatic order and that this order could easily persist in infinite systems.
Once the physics of the infinite system is realised, the discrepancies
between our simulations and many others could probably be easily
resolved.

\section{Acknowledgements}
We wish to thank V. M. Vinokur for some very interesting
and helpful discussions relating to the work and to his paper
in ref. \cite{koshvino}. One of us (SS) would also like to acknowledge
the EPSRC and DRA Malvern for financial support and the other (HJJ)
was supported by the EPSRC under grant no. Gr/J 36952.

$\dagger$ Address from 15 November 1996: Department of Chemistry,
The University of Cambridge, Lensfield Road, Cambridge CB2 1EW, UK

\end{multicols}
\end{document}